\documentclass[
  ,final            
  ,numberedheadings 
  ]
  {aipproc}

\layoutstyle{8x11single}

\begin{document}

\title{$W$ \& $Z$ Production and Asymmetries at the Tevatron}




\author{D.~S.~Waters \\ {\normalsize(for the CDF \& D\O\ collaborations)}}{
  address={Department of Physics \& Astronomy, University College London, London WC1E 6BT, U.K.}
}

\begin{abstract}
An overview of $W$ and $Z$ production in high energy hadron collisions is given. $W$ and $Z$ cross section
and asymmetry measurements from CDF and D\O\  are described, with particular emphasis on recent results. The
current status of precision $W$ mass and width measurements is reported. The fundamental physics parameters that 
can be extracted from these measurements are described, and the relevance of $W$ and $Z$ production studies for
the LHC is pointed out.
\end{abstract}

\maketitle


\section{Introduction}

$W$ and $Z$ bosons are produced in abundance in high energy hadron collisions. Their large mass scale and 
well known fermion couplings make their production cross sections accurately calculable in perturbative QCD, 
and their leptonic decay modes provide distinctive experimental signatures that are relatively easy to
identify, trigger on and separate from large backgrounds. In many respects vector boson production has 
become one of the most important ``standard candles'' in experimental hadron collider physics. At the 
Tevatron, measurements of $W$ and $Z$ inclusive cross sections are routinely used to validate detector and
trigger performance and stability. $Z\to l^{+}l^{-}$ events in particular, due to their negligible 
backgrounds and trigger redundancy as well as precisely known mass, are the most important samples for 
the experimental determination of energy and momentum scales and lepton identification efficiencies. 
Inclusive vector boson cross section measurements are necessary starting points for more detailed
measurements, for example differential cross section measurements, exclusive measurements (e.g. $W+n-{\rm jet}$),
measurements of rare processes (e.g. diboson production) and precision measurements of vector boson
properties. At the Large Hadron Collider (LHC) the use of massive vector bosons as standard candles may be 
taken one step further with their yields being used to either compute, or provide an alternative 
definition of, collider integrated luminosities.

In the following sections the basic physics of vector boson production at hadron colliders is reviewed, 
and the relevant parameters of the CDF and D\O\ detectors are stated. Measurements of inclusive and
differential production cross sections are discussed in section~\ref{sec:xs}, as well as some of the 
particle properties that can be extracted from these measurements. Asymmetry measurements in $Z$
and $W$ production are presented in section~\ref{sec:asym}. Finally, the status of precision direct 
measurements of the $W$ boson mass and width at the Tevatron is given in section~\ref{sec:precision}.

\section{Massive Vector Boson Production at Hadron Colliders}

\begin{figure}
  \includegraphics[height=.175\textheight]{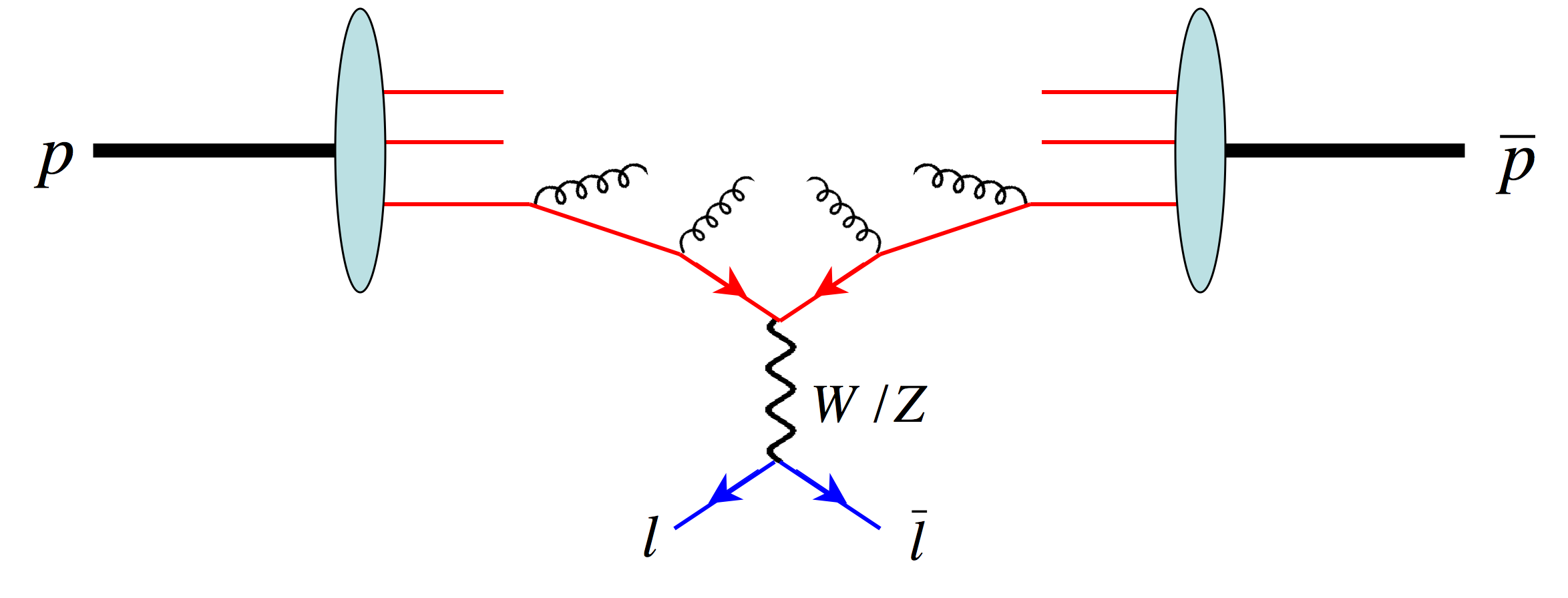}
  \caption{A schematic of vector boson production in high energy proton-antiproton collisions. Energetic 
  quark-antiquark pairs annihilate to produce a $W$ or $Z$ boson, which subsequently decays, in this
  case to a lepton-antilepton pair. QCD effects need to be taken into account to give a full description of
  vector boson production kinematics, as described in the text.}
\label{fig:w_z_production_schematic}
\end{figure}

Figure~\ref{fig:w_z_production_schematic} indicates the phenomenological ingredients required for a 
complete description of vector boson production in hadron collisions. At leading order, a quark-antiquark
pair annihilate to produce a $W$ or $Z$ boson, which subsequently decays to a fermion-antifermion pair.
Cross sections are computed as a convolution of partial cross sections (${\rm d}\sigma_{q\bar{q}\to W/Z\to l\bar{l}}$) 
over parton distribution functions (PDF's; $f_{i}(x)$, where $x$ is the fraction of the proton's momentum carried by
quark flavor~$i$):
\begin{displaymath}
 \sigma_{p\bar{p}\to W/Z\to l\bar{l}}=\int{\sum_{i,j=u,d,s (,c,b)}{[f_{i}^{q}(x_{p})f_{j}^{\bar{q}}(x_{\bar{p}}) + f_{i}^{\bar{q}}(x_{p})f_{j}^{q}(x_{\bar{p}})}] 
 \times  {\rm d}\sigma_{q\bar{q}\to W/Z\to l\bar{l}} \;\;\; {\rm d}x_{p}\; {\rm d}x_{\bar{p}}}\;\; .
\end{displaymath}
The longitudinal momentum (or rapidity) distribution of the produced vector bosons depends directly on 
the PDF's. The rate and kinematic distributions of the decay fermions additionally depend on the couplings
and branching ratios, as well as the mass ($M_{W/Z}$) and width ($\Gamma_{W/Z}$), of the produced vector boson:
\begin{displaymath}
 {\rm d}\sigma_{q\bar{q}\to W/Z\to l\bar{l}}(\hat{s},\theta_{l},\phi_{l}) \; \propto \; {\rm couplings} \; \times \; \left[ \frac{1}{(\hat{s}-M^{2}_{W/Z})^{2} + (\Gamma_{W/Z}\hat{s}/M_{W/Z})^{2}} \right] \;\; ,
\end{displaymath}
where the parton-parton center of mass energy $\sqrt{\hat{s}}\,=\,E_{CM}\,\sqrt{x_{p}x_{\bar{p}}}$.

QCD corrections to this process can be divided into perturbative and non-perturbative parts. Perturbative 
corrections, either in the form of fixed higher order matrix elements or parton showers, modify the vector 
boson production kinematics - most notably the vector boson transverse momentum ($p_{T}$) distributions - 
and give rise to final states with multiple, sometimes high-$p_{T}$ partons. Non-perturbative 
effects, sometimes understood as intrinsic-$k_{T}$ distributions for the colliding partons, are especially 
important for obtaining a complete description of the low end of the vector boson $p_{T}$ spectra. Finally, 
QCD corrections modify the helicity and therefore decay angular distributions of the vector bosons, a 
small but measurable effect that is important to take into account for precision measurements.
%
%

By far the most important electroweak corrections to vector boson production and decay are those due to 
final state photon radiation from charged leptons, where the effect on lepton kinematics and identification
is appreciable. Initial state radiation from the colliding quarks and, in the case of $W$ production, radiation 
from the $W$ itself, are important to consider in analyses of high-$p_{T}$ photons produced in association 
with a vector boson, but have a negligible impact on more inclusive $W$ and $Z$ measurements. 

%
%

\section{The CDF and D\O\ Experiments at the Tevatron}

CDF and D\O\ are general purpose $\sim 4\pi$ detectors, efficiently triggering on and identifying leptonic
decays of $W$ and $Z$ bosons over significant regions of phase space~\cite{cdf_detector,d0_detector}. 
A particular strength of the CDF detector is the excellent central tracking resolution 
$\delta(p_{T})/p_{T} \approx 0.0005 \times p_{T} \; ({\rm GeV/c})$ [$|\eta|<1$, beam constrained]. The central
electromagnetic calorimeter has an energy resolution $\delta(E_{T})/E_{T} \approx 13.5\%/\sqrt{E_{T}} \oplus 1.5\%$.
Muon chambers cover the pseudorapidity region $|\eta|<1$ and forward calorimeters cover the region out to
$|\eta|<3.6$, important for the accurate reconstruction of missing-$E_{T}$. D\O\ has significantly larger muon 
acceptance with chamber coverage and accurate momentum determination out to $|\eta|<2$, and 
excellent hermeticity with calorimeter coverage out to $|\eta|<4.2$.

The performance of the Tevatron collider continues to increase, such that at the time of this meeting 
is has delivered almost $1.5$~${\rm fb^{-1}}$ to each experiment. The measurements presented here 
are based on the first $\sim 400$~${\rm pb^{-1}}$ of Run~II data.

\section{Vector Boson Production Cross Sections}\label{sec:xs}

\begin{figure}
$\begin{array}{c@{\hspace{.5in}}c}
  \includegraphics[height=.35\textwidth]{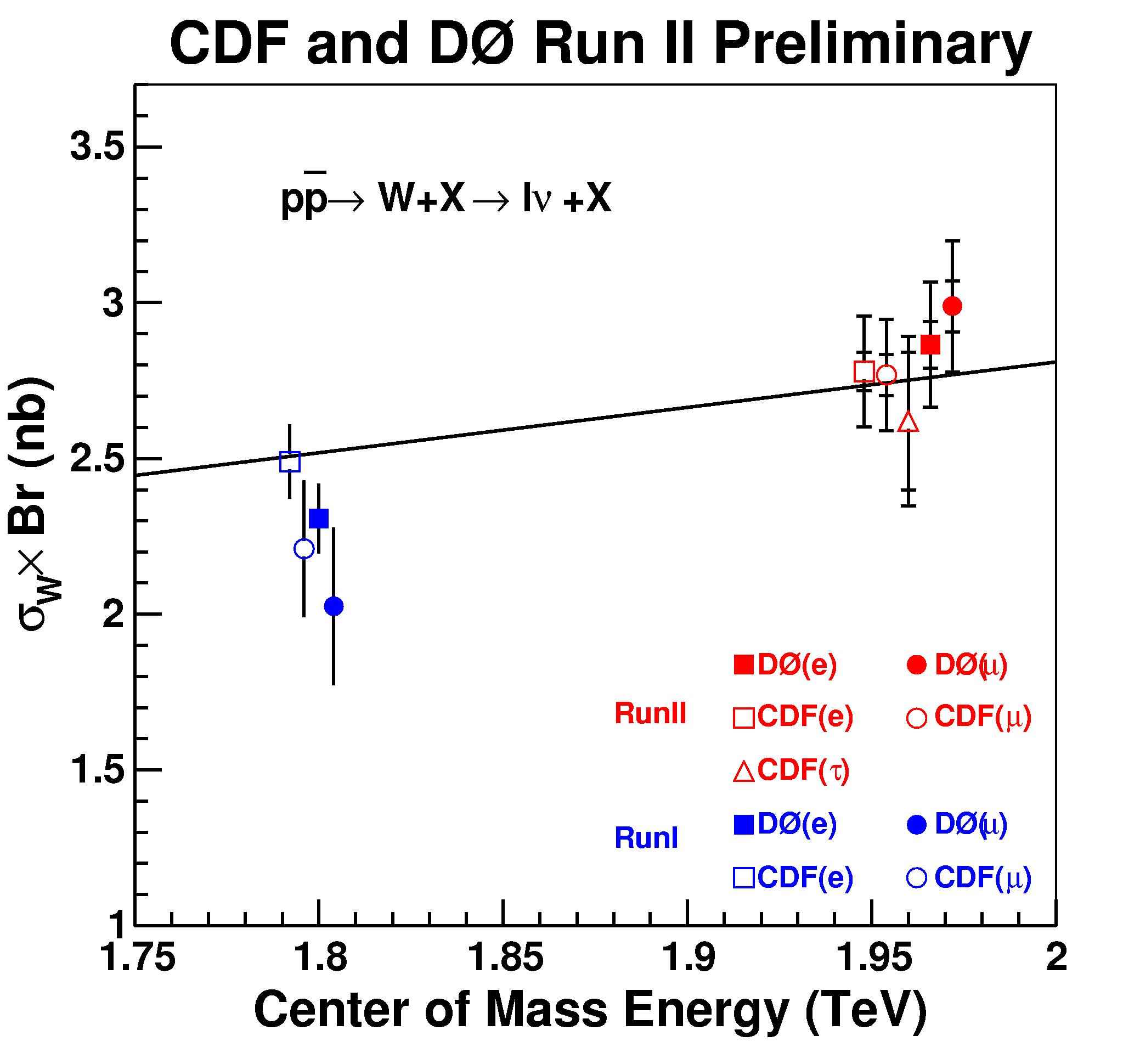}  &
  \includegraphics[height=.35\textwidth]{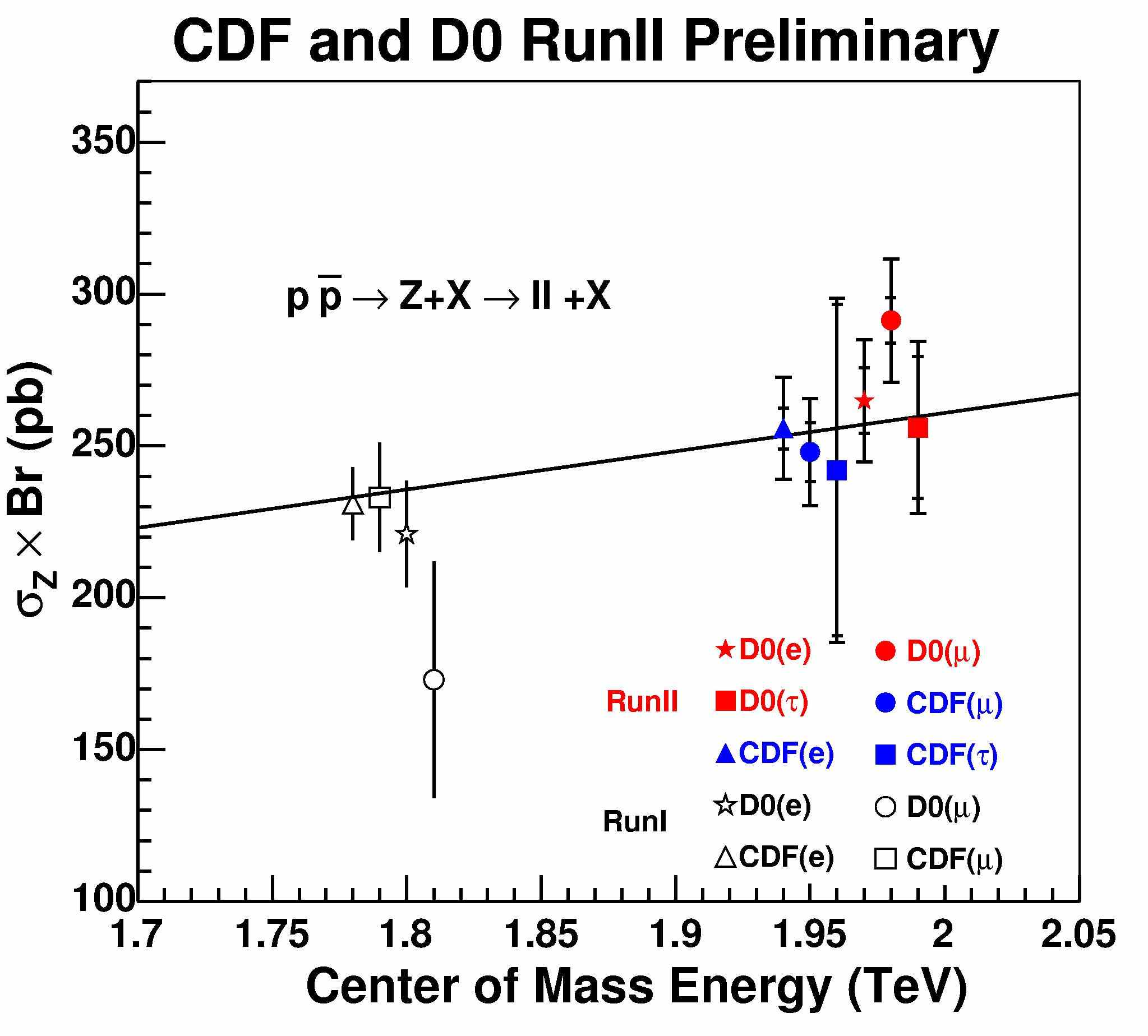} \\ [0.0 cm]
\end{array}$
\caption{A compilation of Run~I and Run~II Tevatron results on the measurement of {\bf (left)} $\sigma_{W}\times{\rm BR}(W\to l\nu)$
and {\bf (right)} $\sigma_{Z}\times{\rm BR}(Z\to l^{+}l^{-})$. The measurements are compared to NNLO predictions; see text for 
further details.}
\label{fig:xsec_w_z_inclusive}
\end{figure}

The measurements of inclusive $W$ and $Z$ production cross-sections in leptonic decay modes 
provide a benchmark for all analyses of events containing high $p_{T}$ leptons at the Tevatron. A compilation
of both Run~I and Run~II measurements from CDF and D\O, based on Run~II datasets between $72$~${\rm pb^{-1}}$ and
$350$~${\rm pb^{-1}}$, is shown in figure~\ref{fig:xsec_w_z_inclusive}, where the good agreement with predictions 
calculated at NNLO in QCD can be observed~\cite{cdf_w_z_xsec}. 

In electron and muon decay channels, the best measurements are systematically limited at the $1-2\%$ 
level, not including an overall $6\%$ uncertainty on the integrated luminosity (completely correlated for all
measurements from a single experiment, and partially correlated between CDF and D\O). The largest 
contributions to the non-luminosity systematic derive from PDF uncertainties, and residual experimental
uncertainties on lepton identification efficiencies and backgrounds. 
Interestingly, with non-luminosity systematic uncertainties of $\sim 2\%$, and given the uncertainty on the 
theoretical cross section predictions of approximately $2\%$, the CDF measurements are already providing a 
cross check of the integrated luminosity measured for that experiment~\footnote{Integrated luminosity 
corrections for the D\O\ experiment are currently being re-evaluated.}.

%
The ratio of $W$ to $Z$ leptonic cross sections can be written as:
\begin{displaymath}
R=\frac{\sigma_{W}\times{\rm BR}(W\to l\nu)}{\sigma_{Z}\times{\rm BR}(Z\to l^{+}l^{-})} =
 \frac{\sigma_{W}}{\sigma_{Z}}  \cdot  \frac{\Gamma_{Z}}{\Gamma_{Z\to l^{+}l^{-}}}  \cdot  \frac{\Gamma_{W\to l\nu}}{\Gamma_{W}} \;\;.
\end{displaymath}
With the ratio of inclusive production cross sections taken from a NNLO calculation, the measurement of the $Z$ leptonic branching
ratio from LEP and a Standard Model calculation of the $W$ leptonic decay partial width, a measurement of $R$ can therefore
be interpreted as an indirect determination of the full $W$ decay width. The CDF $72$~${\rm pb^{-1}}$ analysis results in a 
measurement $\Gamma_{W}=2.092 \pm 0.042$~${\rm GeV}$ that is already comparable in precision with the world average result.
Analyses are currently being developed with the goal of minimizing the systematic uncertainty on the cross section ratio rather than
the individual $W$ and $Z$ cross sections, in order to further improve the indirect $W$ width determination.

\begin{figure}
$\begin{array}{c@{\hspace{.5in}}c}
  \includegraphics[height=.35\textwidth]{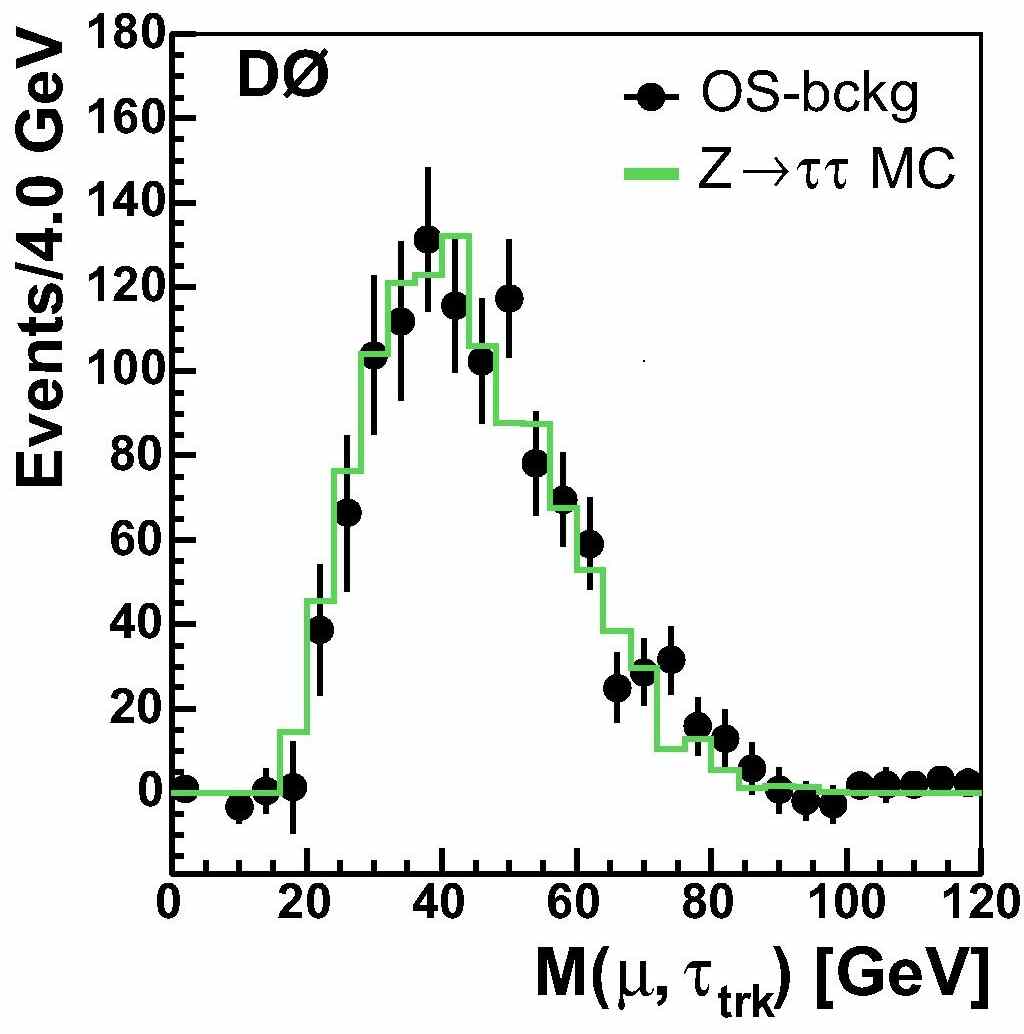}  &
  \includegraphics[height=.35\textwidth]{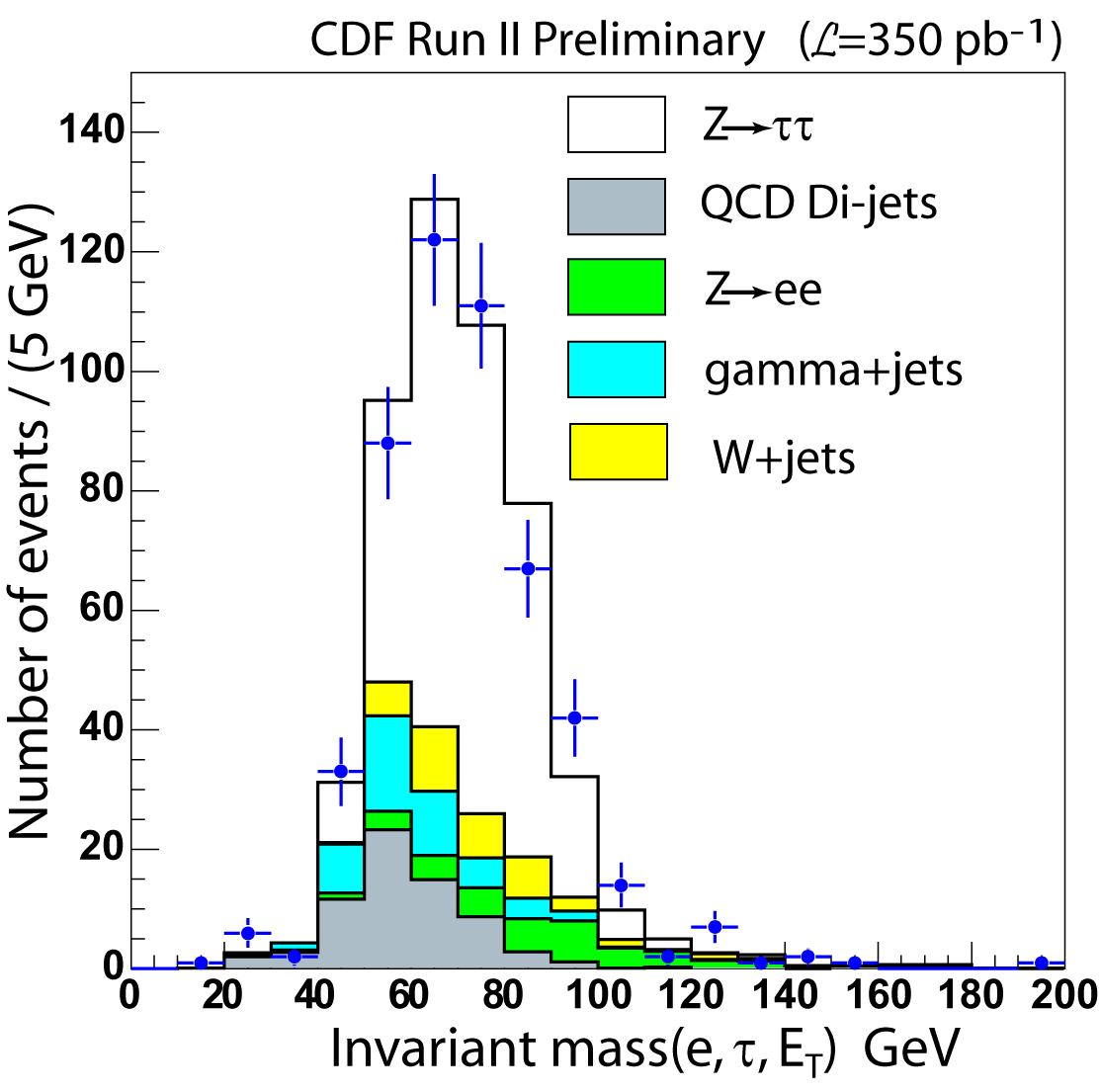} \\ [0.0 cm]
\end{array}$
\caption{{\bf (left)} The invariant mass distribution for the muon and tau-track(s) for a D\O\ analysis of 
$p\bar{p}\to Z (+X) \to \tau(\mu)\tau(h/e)$; {\bf (right)} the visible mass for a CDF analysis of 
$p\bar{p}\to Z (+X) \to \tau(e)\tau(h)$. See text for further details.}
\label{fig:z_tt}
\end{figure}

Tau decay modes of $W$ and $Z$ bosons are experimentally much more challenging in many respects,
including triggering, identification and separation from backgrounds. However the physics reward is
significant, allowing tests of $3^{\rm rd}$ generation lepton universality and establishing a benchmark for many 
searches, most notably the search for the MSSM Higgs bosons. Both CDF and D\O\ have recently performed
cross section measurements for $Z\to\tau^{+}\tau^{-}$ production, where one $\tau$ decays leptonically
and the other hadronically~\cite{z_tt}. The reconstruction of the hadronic tau decay modes involves combining
information from both the tracking and calorimetry, with constraints from the tau mass and decay topologies.
Figure~\ref{fig:z_tt} shows the mass distributions from the two analyses, showing good agreement with Monte
Carlo after applying background corrections. The corresponding cross sections:
\begin{center}
$\sigma_{Z}\times{\rm BR}(Z\to\tau^{+}\tau^{-}) = 237 \pm 15 \;  {\rm (stat.)} \pm 18 \;  {\rm (syst.)} \pm 15 \;  {\rm (lum.)} \;\; {\rm pb} \;\;$ [D\O ; $226$~${\rm pb^{-1}}$]
$\sigma_{Z}\times{\rm BR}(Z\to\tau^{+}\tau^{-}) = 265 \pm 20 \;  {\rm (stat.)} \pm 21 \;  {\rm (syst.)} \pm 15 \;  {\rm (lum.)}  \;\;{\rm pb} \;\;$  [CDF; $350$~${\rm pb^{-1}}$]
\end{center}
agree very well with measurements in the electron and muon channels.
In addition CDF have a measurement of the $W$ production cross section in the hadronic tau decay channel using
$72$~${\rm pb^{-1}}$, and a corresponding determination of the ratio 
$BR(W\to\tau\nu)/BR(W\to e\nu)=0.99 \pm 0.04 \; {\rm (stat)} \pm 0.07 \; {\rm (syst)}$. An improved measurement of
this ratio with much higher integrated luminosity will be particularly interesting to compare to measurements
from LEP~II.

\subsection{Differential Cross Sections}

\begin{figure}
$\begin{array}{c@{\hspace{.45in}}c}
  \includegraphics[height=.32\textwidth]{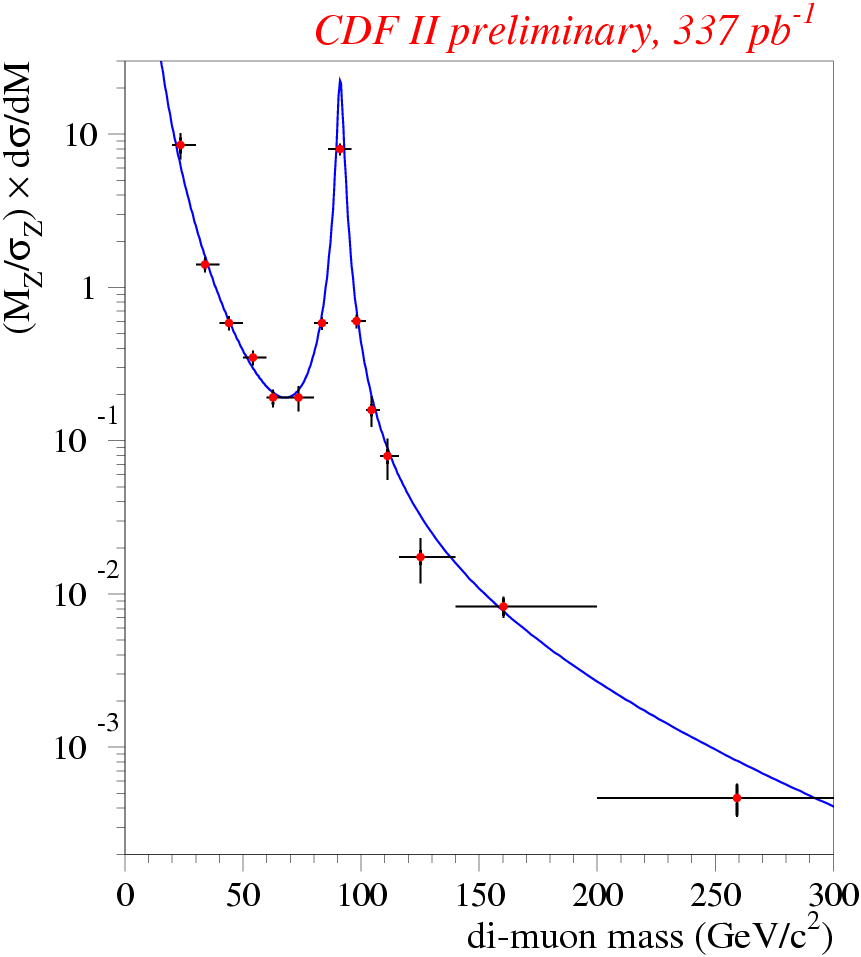}  &
  \includegraphics[height=.32\textwidth]{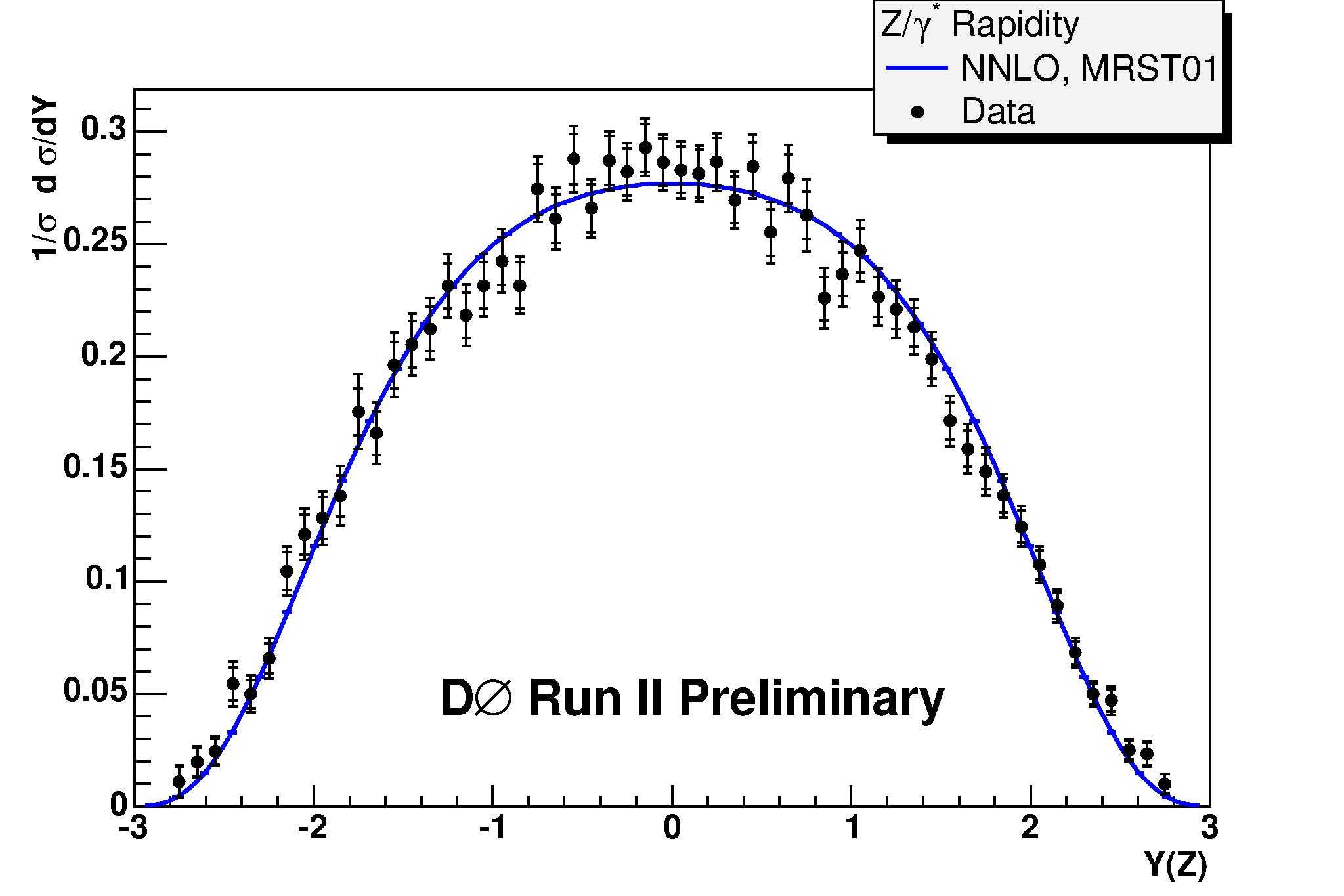}  \\
\end{array}$
\caption{{\bf (left)} The differential cross section for dimuon pairs as a function of dimuon mass from CDF, 
normalized to the $Z$ pole cross section. Note the good agreement between data and theory over four orders of 
magnitude in cross section. {\bf (right)} The unit normalized differential cross section for $Z$ production as a 
function of vector boson rapidity, measured by D\O\ in the electron channel. }
\label{fig:xsec_diff}
\end{figure}

CDF and D\O\ have moved on from inclusive cross sections to the measurement of differential cross sections. 
Measuring the Drell-Yan cross section ${\rm d}\sigma_{l^{+}l^{-}}/{\rm d}M$ over as large a mass range
as possible  controls an important background to searches for new physics in dilepton channels. Both the
mass and the rapidity~\footnote{The rapidity $Y$ is a function of energy and longitudinal momentum:~~
$Y=\frac{1}{2}\ln\left(\frac{E+p_{Z}}{E-p_{Z}}\right)$} differential cross section ${\rm d}\sigma_{l^{+}l^{-}}/{\rm d}Y$ can, with 
sufficient integrated luminosity, be used to constrain PDF's and test higher-order QCD. 
Figure~\ref{fig:xsec_diff} shows measured Drell-Yan mass ($Z$ boson rapidity) differential cross sections in
muon (electron) channels from CDF (D\O), both with $337$~${\rm pb^{-1}}$ of data. The agreement with
theoretical predictions, after careful unfolding of the data for smearing effects, is very good.

Differential cross section measurements for $W$'s are harder than for $Z$ bosons, due to the 
incomplete kinematic reconstruction of events with neutrinos in the final state. As a first attempt, the CDF
experiment has measured the cross section for $W$ production with forward electrons ($1.2\le |\eta_{e}|\le 2.8$)
in $223$~${\rm pb^{-1}}$ of data, using silicon-only small angle tracking and with a different triggering 
strategy to central leptons. The ratio of central to forward $W$ cross sections is particularly interesting:
\begin{displaymath}
R^{central/forward}_{\rm CDF}=0.925\pm0.033\;.
\end{displaymath}
The prediction for this ratio depends on the choice of PDF set:
\begin{displaymath}
R^{central/forward}_{\rm CTEQ~6.1}=0.924\pm0.037 \;\;\; , \;\;\; R^{central/forward}_{\rm MRST~01E}=0.941\pm0.012\;.
\end{displaymath}
Although not yet significant, there is clearly hope that with more data, this and other differential cross section
measurements at the Tevatron can yield useful inputs to global PDF fits.

\section{Asymmetries}
\label{sec:asym}


In addition to cross sections, measured asymmetries are particularly useful for the extraction of underlying physics
parameters such as electroweak couplings and PDF's. Many systematics cancel in the measurement of asymmetries
as opposed to absolute cross sections, and in certain cases (most notably the search for $Z'$ bosons), hints of new or
unexpected physics may first turn up as asymmetry rather than rate anomalies.

\begin{figure}
$\begin{array}{c@{\hspace{.25in}}c}
  \includegraphics[height=.32\textwidth]{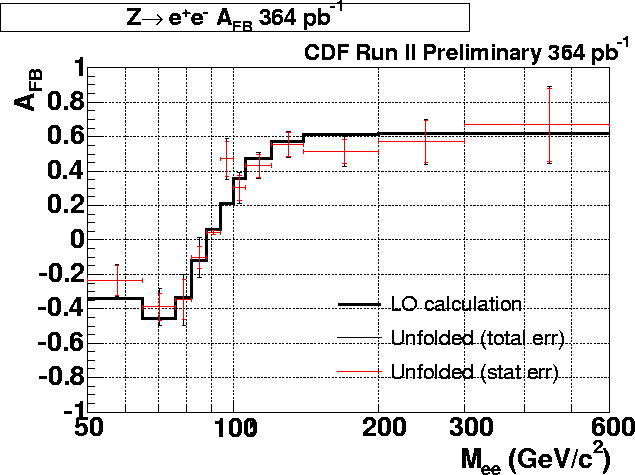} &
  \includegraphics[height=.32\textwidth]{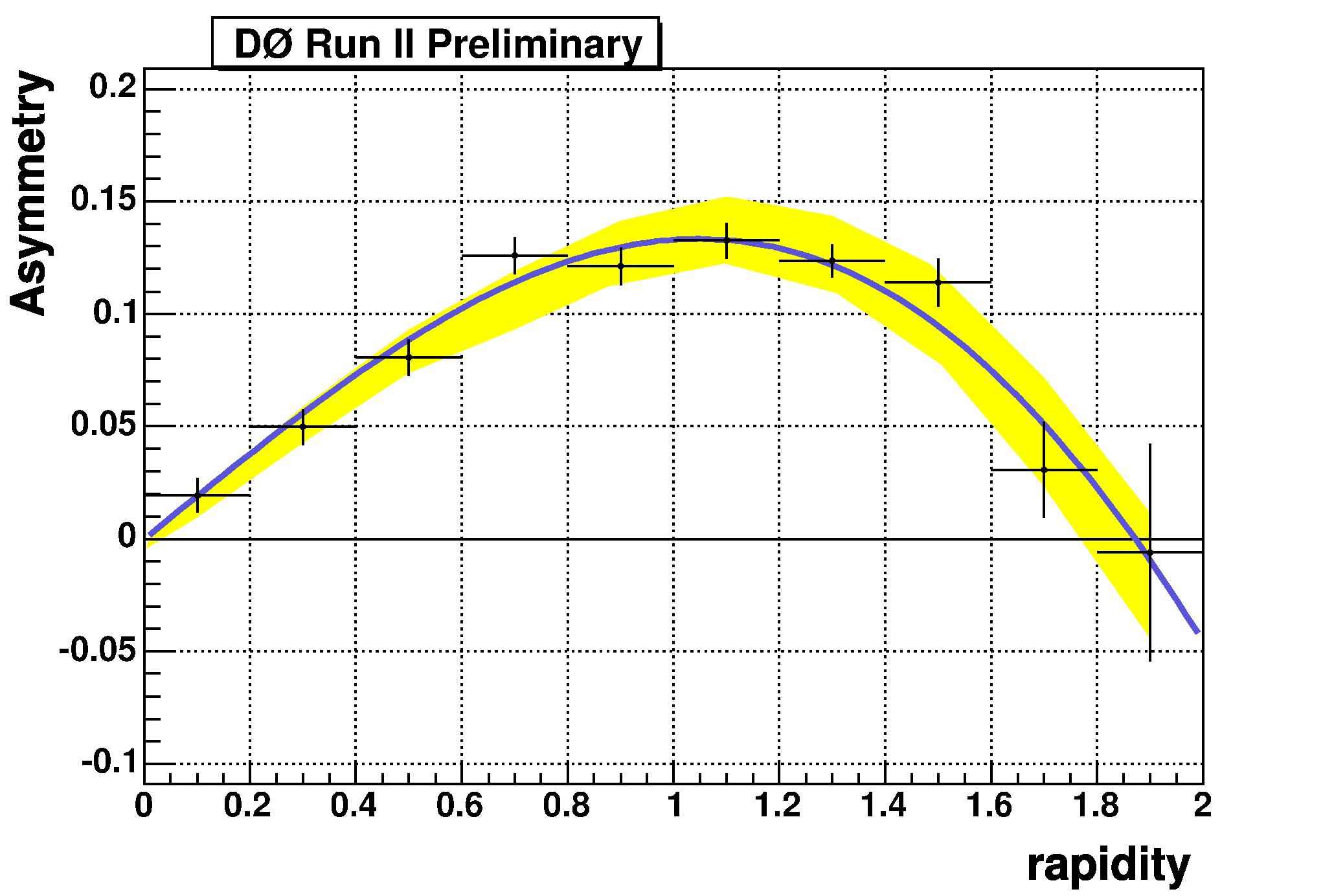} \\ [0.0 cm]
\end{array}$
\caption{{\bf (left)} A measurement of the forward-backward asymmetry as a function of di-electron mass in 
Drell-Yan data at CDF.  {\bf (right)} The $W$ charge asymmetry measured as a function of the muon 
pseudorapidity in $230$~${\rm pb^{-1}}$ of D\O\ data. The blue curve is the expectation using the
MRST-02 PDF set, while the yellow band is the range of predictions using the CTEQ6.1M error PDF sets. 
See text for further details.}
\label{fig:asym}
\end{figure}

The Drell-Yan forward-backward asymmetry compares the fractions of events for which the negatively charged lepton
is produced in the forward and backward hemispheres, defined in the parton-parton center of mass system:
\begin{displaymath}
A_{FB}=\frac{\sigma_{F}-\sigma_{B}}{\sigma_{F}+\sigma_{B}}\;\; , \;\;\;\; {\rm where} \;\;\;
\sigma_{F(B)}=\int_{0(-1)}^{1(0)}\frac{{\rm d}\sigma}{{\rm d}\cos{\theta^{*}}}{\rm d}\cos{\theta^{*}}\;\;.
\end{displaymath}
The result of a recent CDF update to their published $A_{FB}$ analysis~\cite{cdf_afb} is shown in figure~\ref{fig:asym}. The 
structure around the $Z$ pole, as well as the measured asymmetry at high mass where indications of new physics may first 
be expected to be observed, agree well with theoretical predictions. The previous analysis has also used the measured
asymmetry to place constrains on the neutral current light quark couplings. Although less precise,
the Tevatron $A_{FB}$ data, along with data from the analysis of deep inelastic scattering at HERA, do break a 
degeneracy in the determination of light quark couplings at LEP, providing an important confirmation of the Standard Model.

The observed $W$ charge asymmetry, defined with respect to the decay lepton direction, is a combined effect
of both the underlying $W$ production asymmetry and $(V-A)$ decay asymmetry:
\begin{displaymath}
A(\eta_{l})=\frac{{\rm d}\sigma_{+}/{\rm d}\eta_{l} - {\rm d}\sigma_{-}/{\rm d}\eta_{l}}{{\rm d}\sigma_{+}/{\rm d}\eta_{l} + {\rm d}\sigma_{-}/{\rm d}\eta_{l}}=A(y_{W})\otimes (V-A)\;\;.
\end{displaymath}
The expected $W$ charge asymmetry is sensitive to the assumed PDF, in particular the $(d/u)$ ratio at high-$x$,
and benefits from increased statistics with respect to the $Z$ rapidity measurement. Important experimental issues 
are forward lepton identification and, crucially, well controlled charge mis-identification rates in order to correct
 the observed asymmetry for any dilution due to lepton charge mis-assignment.
Figure~\ref{fig:asym} shows a recent $W$ charge asymmetry measurement by D\O\ in the muon channel, where
charge mis-identification rates were kept at the $10^{-4}$ level out to muon pseudorapidities of 2. Interestingly,
the measurement uncertainties are now appreciably smaller than the variation in theoretical predictions that
can be generated by varying the input PDF set. This indicates that this data would already contribute additional 
constraining power to future global PDF fits.
CDF have  measured the asymmetry using electrons in a smaller dataset~\cite{cdf_charge_asym}. They have 
additionally shown that certain lepton kinematic regions are more sensitive to PDF variations than others.
Analyses are currently underway to fully exploit all the kinematic information available in $W$ events with the
goal of unfolding directly back to the underlying $W$ production asymmetry.

\section{Precision Measurements of the $W$ Mass and Width}
\label{sec:precision}

The precision determination of the $W$ mass is one of the most important measurements to be performed
at the Tevatron. $M_{W}$ is a critical input to Standard Model fits which can constrain the mass of the 
unobserved Higgs boson or, subsequent to a discovery of a Higgs boson, may give indications as to what  
lies beyond the Standard Model. The width of the $W$ is a less sensitive observable in global electroweak fits,
but the direct measurement of the $W$ width nevertheless confirms a basic prediction of the Standard Model
and provides a useful cross check of indirect measurements.

The most precise top mass measurements at the Tevatron are at the level of $1\%$. By contrast the 
goal for the $M_{W}$ is a measurement substantially better than $0.1\%$. This highlights the 
extremely difficult nature of this measurement, requiring exquisite understanding of $W$ production
and decay, detector response and the effect of backgrounds. CDF have estimated that in the first
$200$~${\rm pb^{-1}}$ of Run~II data, they will measure the $W$ mass with a precision of at least
$76$~${\rm MeV}$, better than the combined Tevatron Run~I uncertainty. D\O\ have started with a 
direct measurement of the $W$ width in $177$~${\rm pb^{-1}}$ of Run~II data: 
$\Gamma_{W} = 2.011 \pm 0.093 \;  {\rm (stat.)} \pm 0.107 \;  {\rm (syst.)}$~${\rm GeV}$.
Both experiments are finalizing new measurements of the $W$ mass, with an ultimate goal for Run~II
of $30-40$~${\rm MeV}$ per experiment.

\section{Conclusions}

The CDF and D\O\ experiments have completed benchmark measurements of inclusive $W$ and $Z$ production,
as well as differential cross section and asymmetry measurements. These measurements are providing important
information on fundamental $W$ and $Z$ properties, as well as parton distribution functions and other aspects of vector 
boson production physics. With the basic signatures and experimental techniques involved in the measurement of 
$W$ and $Z$ production well understood, attention is now focused on optimizing analyses for sensitivity to PDF's and
other physics parameters of interest. All of this information is being brought to bear in the precision measurement of
the $W$ mass in Run~II, the first results on which are expected soon.

Ultimately, measurements of $W$ and $Z$ production at the Tevatron will be used as inputs to the
accurate modeling of vector boson production at the LHC, crucial for using these processes
to help commission and calibrate LHC detectors. In general, vector boson production promises to 
continue to be one of the most useful standard candles of hadron collider physics.





\bibliographystyle{aipproc}   


\begin{thebibliography}{9}

\bibitem{cdf_detector} D.~Acosta {\em et al.} (CDF Collaboration), Phys.~Rev.~D {\bf 71}, 032001 (2005).
\bibitem{d0_detector} V.~M.~Abazov {\em et al.} (D\O\ Collaboration), "The Upgraded D\O\ Detector", 
hep-physics/0507191 (submitted to Nucl.~Instrum.~Methods~Phys.~Res.~A).
\bibitem{cdf_w_z_xsec} D.~Acosta {\em et al.} (CDF Collaboration), Phys.~Rev.~Lett. {\bf 94}, 091803 (2005); 
hep-ex/0508029 (submitted to Phys.~Rev.~D).
\bibitem{z_tt} V.~M.~Abazov {\em et al.} (D\O\ Collaboration), Phys.~Rev.~D {\bf 71}, 072004 (2005);
A.~Safonov (for the CDF Collaboration), Nucl.~Phys.~Proc.~Suppl. {\bf 144} 323 (2005); 
\bibitem{cdf_afb} D.~Acosta {\em et al.} (CDF Collaboration), Phys.~Rev.~D {\bf 71}, 052002 (2005).
\bibitem{cdf_charge_asym} D.~Acosta {\em et al.} (CDF Collaboration), Phys.~Rev.~D {\bf 71}, 051104(R) (2005).

\end{thebibliography}

\IfFileExists{\jobname.bbl}{}
 {\typeout{}
  \typeout{******************************************}
  \typeout{** Please run "bibtex \jobname" to optain}
  \typeout{** the bibliography and then re-run LaTeX}
  \typeout{** twice to fix the references!}
  \typeout{******************************************}
  \typeout{}
 }

\end{document}